\documentclass[aps,pre,twocolumn,superscriptaddress,showpacs,draft]{revtex4}
\usepackage{amsmath}
\usepackage{amsfonts}
\usepackage{amssymb}
\input{epsf}

\newcommand{\rem}[1]{}

\newcommand{\ps}{\ps}

\begin{document}

\title{Transport phenomena in the asymmetric quantum multibaker map}

\author{Leonardo Ermann}
\affiliation{Departamento de F\'\i sica, CNEA, Libertador 8250, (C1429BNP) Buenos Aires, Argentina}
\affiliation{Departamento de F\'\i sica, FCEyN, UBA, Pabell\'on 1 Ciudad 
Universitaria, C1428EGA Buenos Aires, Argentina}

\author{Gabriel G. Carlo} 
\affiliation{Departamento de F\'\i sica, CNEA, Libertador 8250, (C1429BNP) Buenos Aires, Argentina}
\author{Marcos Saraceno} 
\affiliation{Departamento de F\'\i sica, CNEA, Libertador 8250, (C1429BNP) Buenos Aires, Argentina}
\affiliation{Escuela de Ciencia y Tecnolog\'\i a, UNSAM, Alem 3901, B1653HIM Villa Ballester, Argentina}

\date{\today}

\pacs{05.45.Mt, 05.40.Jc, 05.60.-k}

\begin{abstract}
By studying a modified (unbiased) quantum multibaker map, we were able to obtain a 
{\em finite} asymptotic quantum current without a classical analogue. 
This result suggests a general method for the design of {\em purely} quantum 
ratchets, and sheds light on the investigation of the mechanisms leading 
to net transport generation by breaking symmetries of quantum systems. 
Moreover, we propose the multibaker map as a resource to study 
directed transport phenomena in chaotic systems without bias. 
In fact, this is a paradigmatic model in classical and quantum 
chaos, but also in statistical mechanics.
\end{abstract}

\maketitle

\section{Introduction}

In recent years many works have investigated different kinds of 
transport phenomena in periodic dynamical systems having no external net 
force or bias (the so-called ratchet effect) \cite{Review}. 
This interest is of fundamental character, but it is also motivated 
by the fact that many possible applications exist. For example, they can 
be useful to understand and develop rectifiers, pumps,
particle separation devices, molecular switches, and transistors.
Also, there is great interest in biology, since the working
principles of molecular motors can be explained in
terms of ratchet mechanisms \cite{Julicher}.
Finally, we would like to mention cold atoms and Bose-Einstein 
condensates as promising fields of application, thanks to recent 
developments of the techniques needed to manipulate them 
\cite{Creffield, PolettiR}. In the following we will have in mind 
systems that present chaotic features, since ratchets generally 
behave this way \cite{CAttractors,Carlo1,Flach1}. Hence, 
methods from classical and quantum chaos become 
extremely useful.      

The explanation of the appearance of a net current, i.e., 
average momentum different from zero, is one of the main 
topics of this research.
In a classical context, this behavior can be 
understood in terms of broken symmetries. 
One of the most convincing points of view up to 
now, is that all symmetries of the system leading to momentum inversion 
(i.e., sign change) should be broken in order to have a 
net directed current \cite{Flach1}. This amounts to say that, being 
only a necessary condition, we can follow Curie's 
principle, and assume that if the current is not forbidden by symmetries, 
then it should be present. 
In non-Hamiltonian cases, the asymmetrization of a chaotic attractor 
leads to a net directed current \cite{CAttractors}. In Hamiltonian 
systems the asymmetrization of a chaotic layer embedded in a mixed 
phase space, for example, has the same consequences for a set 
of initial conditions inside of it \cite{Schanz,DenisovOrig}. 
In this case, a mixed dynamics is in general a necessary 
condition (for a notable 
exception see \cite{quantumTheo,Carlo2}). 

The vast majority of the previously mentioned papers were focused on the classical 
aspects, leaving the quantum side less explored. However, there 
has been several recent publications that deal with this 
part of the problem. These works regard both 
the experimental \cite{quantumExp} (systems of cold atoms, mainly) 
and the theoretical sides \cite{Schanz,quantumTheo,quantumTheo2}. 
In general the quantum versions share symmetry
aspects with their classical counterparts, showing the corresponding
current. But sometimes the relation between 
symmetries and the generated current are less obvious in the quantum case. 
Tunneling, for example, can modify the direction of the current 
\cite{tunneling}.
Interference, in any of its forms, generates more complex behaviors
\cite{Brumer}.  
In fact, we will see that the quantum and classical behaviors can 
be very different.

A remarkable situation appears in some cases, when one can find 
a net quantum current that does not have a classical counterpart. 
This was essentially studied in Hamiltonian (non dissipative) 
systems.
The first time this phenomenon was found was in the modified 
kicked rotor (KR) at quantum resonance (i.e., the usual KR with a  
second harmonic in the kick) \cite{Lund,Poletti}. This is a 
classically chaotic system where the time reversal symmetry changing 
the sign of the momentum is kept. 
Later, in the case of the modified kicked Harper model \cite{Gong}, the classical 
current was found to be exceedingly small in comparison with the quantum one. 
In all of these cases the current does not reach an asymptotic value, 
and in fact these systems were called quantum ratchet accelerators.
We present here a different behavior by means of 
a modified multibaker map \cite{gaspard,wojcik}. 
This system models a particle evolving with free flights and 
collisions with fixed scatterers.
We were able to find a {\em finite asymptotic} 
current that is only present in the quantized version. 
This suggests a general method for 
obtaining {\em purely} quantum ratchets (no classical current), 
without unbounded acceleration. 
We propose this system as a model to study general 
quantum transport phenomena in the presence of asymmetries.

\section{The Model}

\begin{figure}[htp]
\centerline{\epsfxsize=6cm\epsffile{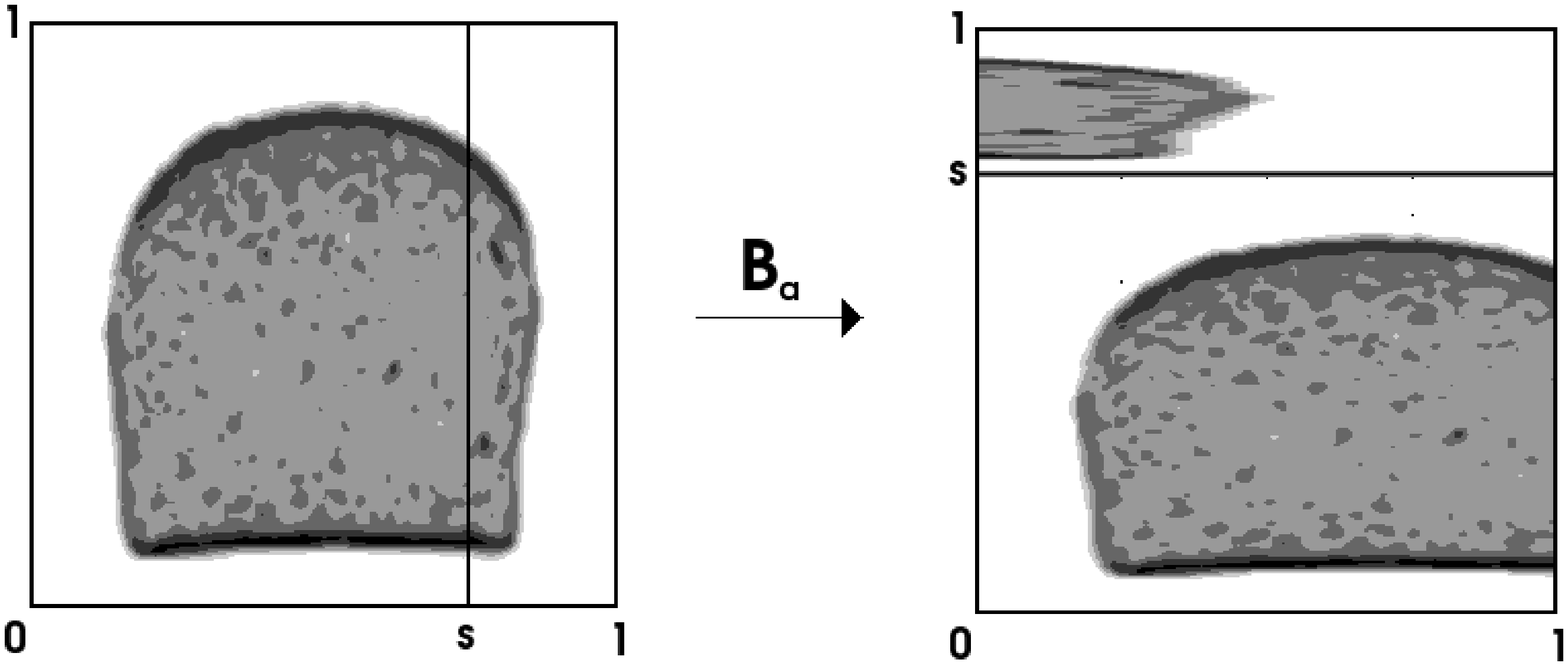}}
\centerline{\epsfxsize=6cm\epsffile{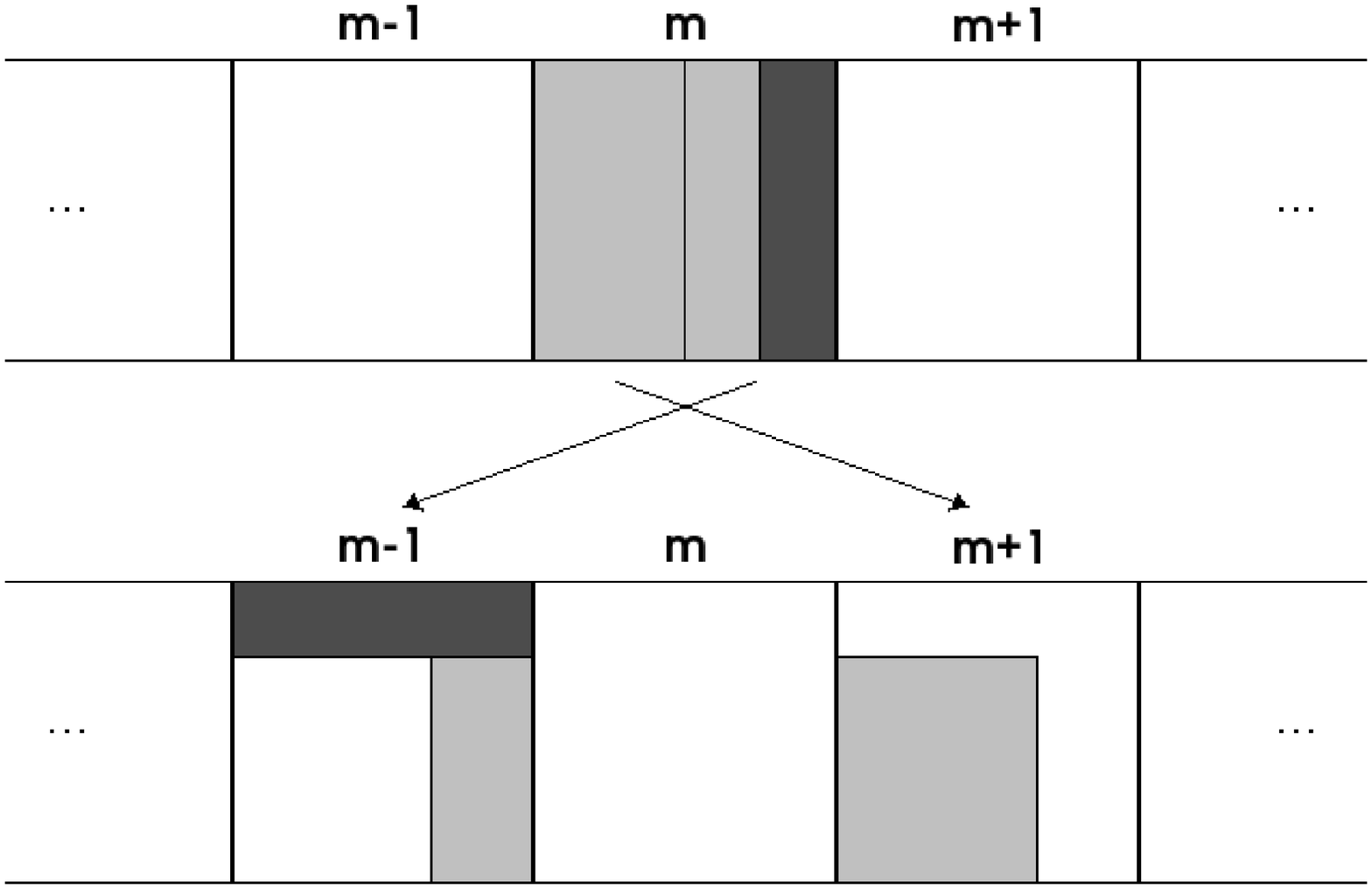}}
\caption{Geometric action of the asymmetric baker's map on 
each cell (top) and of the composition with the translation, 
i.e., the asymmetric multibaker map (bottom).}
\label{fig:bakerasim}
\end{figure}

The well known classical baker's transformation is an area-preserving map 
defined on the unit square phase-space ($0\leq q,p\leq1$). Here we use an 
asymmetric version which divides the phase space into two regions with 
different areas and Lyapunov exponents. The general form of this map can 
be written in terms of one parameter $s\in(0,1)$
\begin{equation}\label{eq:bakerasimGeneral}
 B_{s}(q,p)\equiv\left\lbrace \begin{array}{cc} \left( \frac{1}{s}q, s p\right) 
&0\leq q\leq s\\ 
\left((1-s)^{-1} (q-s),(1-s) p +s \right) &s\leq q\leq 1
\end{array}
 \right. 
\end{equation}
where we can recover the usual symmetrical case by setting $s=1/2$. 
The geometric action of the map is represented in Fig. \ref{fig:bakerasim} (top).

The quantum version of the map is defined in a discrete 
$D$-dimensional Hilbert space with $h=1/D$ in terms of the quantum Fourier 
transform with antiperiodic boundary conditions \cite{voros,saraceno}. 
\begin{eqnarray}
 \hat{B}_{s}&=&\hat{G}^{\dagger}_{D}\left(\begin{array}{cc}\hat{G}_{D_{1}}&0\\0&
\hat{G}_{D_{2}}\end{array}\right)\\
\left(\hat{G}_{D}\right)_{kl}&\equiv& D^{-1/2}e^{-i 2\pi(k+
1/2)(l+1/2)/D} 
\end{eqnarray}
where the allowed values of $s$ are such that $D_{1}=sD$ and $D_{2}=D-D_{1}$ 
are integer numbers. 
The same procedure can be done to obtain an entire family of asymmetric 
quantum baker maps (QBMs) \cite{schackcaves,leo2}. 

The classical multibaker map \cite{gaspard} is defined in a 
two-dimensional lattice where the phase space at each lattice site is a 
unit square. The dynamics of the map is a combination of transport to 
neighboring cells and a local evolution within a cell. The map is defined 
as $M_{s}=B_{s} \circ T$, where the baker term is the 
asymmetric baker map defined in Eq. \ref{eq:bakerasimGeneral} applied on each 
cell $m$, and the transport term is defined as
\begin{equation}
 T=\left\lbrace \begin{array}{cc}\left( m+1,q,p\right) &0\leq q 
\leq 1/2\\
 \left( m-1,q,p\right) &1/2\leq q \leq 1
\end{array}\right. 
\end{equation}
The geometric action of the multibaker map in the phase space of a lattice 
of squares is represented in Fig. \ref{fig:bakerasim} (bottom). 
While the baker's map is asymmetric, the transport term is unbiased as 
phase space volume is transported symmetrically, as can be clearly 
seen. Notice that the transport is entirely due to translations and 
therefore there are no tunneling effects from cell to cell. 

This transformation is not biased neither in the $p$ nor in the $q$ 
coordinate; the reason is that the baker 
transformation maps the unit square onto itself and the 
transport term is balanced, as previously explained.
This is the equivalent to the zero average net force typical 
of the dynamical systems addressed in studies of directed transport \cite{Review}.  
In what follows we will focus on the coarse grained 
current, that in the classical case is defined by 
\begin{equation}
J_{class}(t)=\langle m(t)\rangle-\langle m(t-1)\rangle
\end{equation}
In this expression 
$\langle m(t)\rangle$ is the average value of the cell position $m$ 
for a given ensemble of initial conditions, at a time $t$.
It is easy to see that this definition does not take care 
of the fluctuations inside of each cell. However, we underline that 
there is no bias, making this model completely general. 

From the classical point of view, 
the presence or absence of an asymptotic current follows the general 
criteria specified in \cite{Flach1} and depends on the symmetries 
that reverse the sign of the transport. Here there are two such 
symmetries 
\begin{eqnarray}
S_I: \, \, \, & q \rightarrow 1-q; \, \, \, p \rightarrow 1-p
; \, \, \, T \rightarrow T &\\
S_{II}: \, \, \, &q \rightarrow p; 
\, \, \, p \rightarrow q; \, \, \, T \rightarrow T^{-1}; 
\, \, \, t \rightarrow -t&
\end{eqnarray}
The first one maps $B_{s}$ 
to $B_{1-s}$, and therefore is broken unless $s=1/2$. 
$S_{II}$ is a time reversal symmetry and is preserved at all times. 
Thus, according to the criteria of \cite{Flach1} there cannot be 
an asymptotic classical current for unbiased initial conditions. 
Transient effects can be present for biased conditions but will 
die off very rapidly due to the exponential mixing property of 
the Baker map. Numerical calculations confirm this expectation 
(see inset of Fig. 2(b) and caption).

The quantum multibaker map is the 
composition of a translation on the lattice 
site that depends on the value of the projectors acting on the right 
and left subspaces of the baker map at each cell \cite{wojcik}. This can be written as
\begin{equation}
\label{eq:Qmultibak}
\hat{M}_{s} \equiv \hat{B}_{s} \circ \hat{T}= 
\left( \hat{I} \otimes \hat{B}_{s} \right)
\left( \hat{U} \otimes\hat{P}_{R} + 
\hat{U}^{\dagger} \otimes \hat{P}_{L}\right),
\end{equation}
where $P_R$ and $P_L$ are the projectors and $\hat{U}$ is a 
unitary translation operator acting on the lattice subspace 
$\hat{U} |m\rangle = |m+1\rangle$ (with 
$\{|m\rangle, m=\ldots,-2,-1,0,1,2,\ldots\}$ the basis 
set of the lattice).

\section{Numerical Results}

The discrete time propagation of an initial state 
$\rho_{0}=\rho_{\text{lat}}\otimes\rho_{\text{QBM}}$ is 
\begin{equation}
\rho(t)=\left(\hat{M}_{s} \right)^t\rho_0\left(\hat{M}_{s}^{\dagger} \right)^{t}
\label{qevolution}
\end{equation}
Here we will focus on initial states localized in one site of the 
lattice $\rho_{\text{lat}}=|0\rangle\langle0|$. 
A word regarding the analogy with quantum walks is in order 
here. In fact, this system can be thought as the $D$-dimensional quantum baker 
map coupled to a quantum walker in an infinite dimensional lattice \cite{Brun,leo1}. 
The role of the \emph{quantum coin} in the quantum walk is performed here by the 
quantum baker map. The fact that the \emph{coin} is classically ergodic is 
however an important difference, and is a fundamental reason for having no classical current.


\begin{figure}[tp]
\centerline{\epsfxsize=8.3cm\epsffile{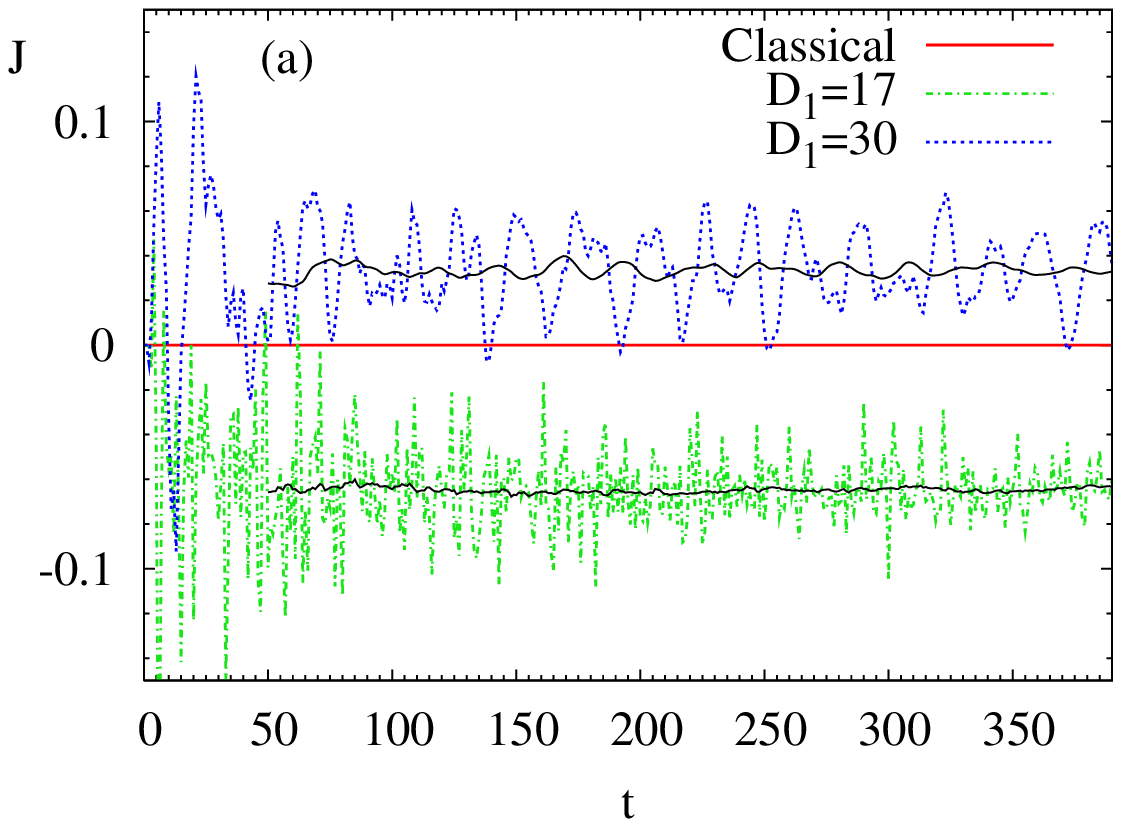}}
\centerline{\epsfxsize=8.3cm\epsffile{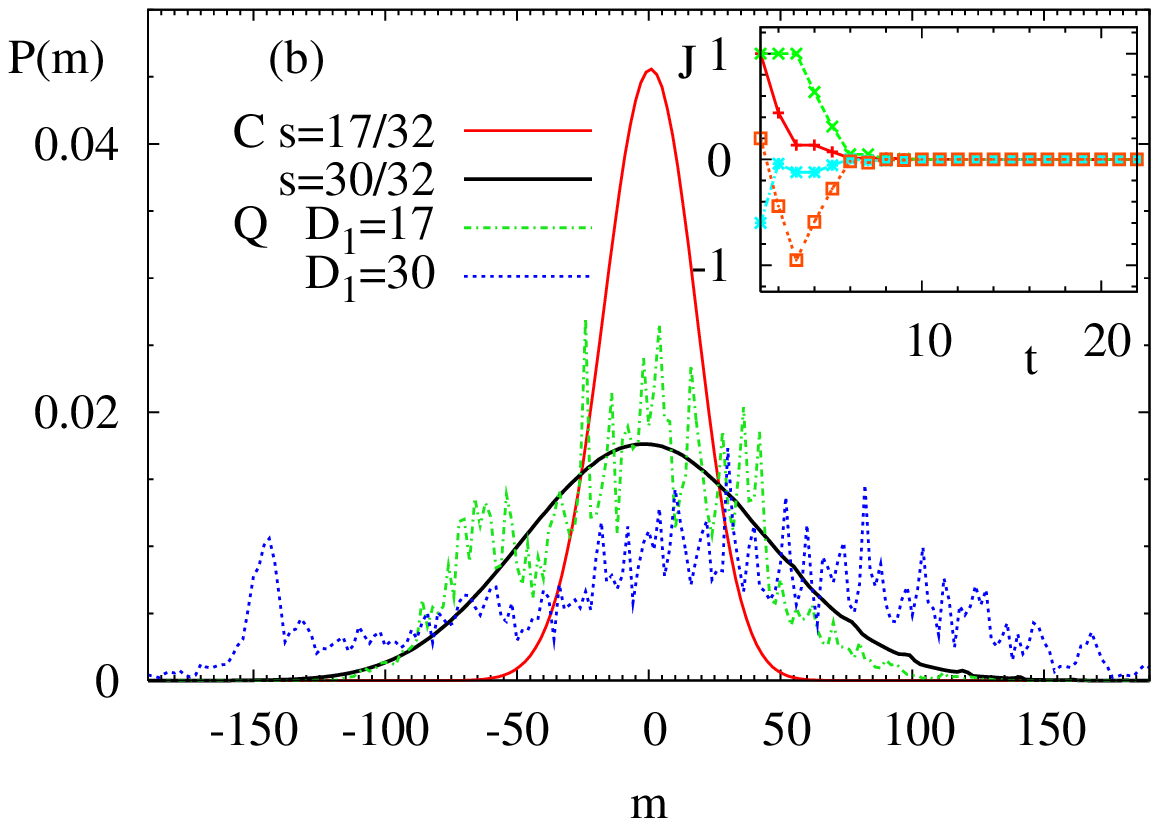}}
\caption{(Color online) (a): Coarse grained current $J$ as a function of time for 
the QBM with $D=32$ and $D_{1}=17$ (dot--dashed line) and $D_{1}=30$ (dotted line). 
In solid lines the smoothed current over 20 steps is shown. The solid line at $J=0$ 
corresponds to the classical current calculated for $10^8$ initial 
conditions. 
(b): The probability distribution $P(m,t)$   
at time $t=200$ for the same cases as before, $s=17/32$ and $s=30/32$ in the classical and quantum version with $D=32$.
In the inset we show how the net current become null in the classical case for different initial conditions, calculated for $10^8$ points in phase space, in particular for two squares of area $1/16$ centered in $(q,p)=(1/8,1/2)$ ($s=0.6$ ($+$) and $s=0.8$ ($\times$)) 
and $(3/8,1/2)$ ($s=0.6$ ($\bigstar$) and $s=0.8$ ($\square$)).
}
\label{fig:Jdet32}
\end{figure}

The transport properties of the system can be computed with the coarse 
grained density of probabilities of the lattice. This distribution is 
obtained by tracing out the internal degrees of freedom inside each cell 
and projecting on the lattice basis. Being 
\begin{equation}
\label{prob}
P(m,t)=Tr\left[\left(\vert m \rangle\langle m\vert \otimes I_{\text{QBM}}\right) \rho(t)\right]
\end{equation}
the mean value of the coarse grained position and the quantum current 
become
\begin{eqnarray}
\langle x(t)\rangle&=&\sum_{m=-\infty}^{\infty} m P(m,t)\\
J(t)&=&\langle x(t)\rangle-\langle x(t-1)\rangle.
\end{eqnarray}

We now turn to analyze the current behavior of this system by means of 
numerical simulations. First, we point out that 
if we take a mixed superposition of centered eigenstates of momentum as the initial 
condition for the QBM, we find that there is no current ($J$) for the symmetric 
case ($s=1/2$). This confirms the need to break the $S_I$ symmetry in order to have 
a net directed transport. Throughout the following 
calculations we consider as initial condition 
the mixed state corresponding to the incoherent superposition of the two central 
momentum eigenstates, i.e. 
\begin{equation}
\rho_{\text{QBM}}=\frac{1}{2}\hat{G}_{D}\left(\left\vert\frac{D}{2}-1
\right\rangle \left\langle \frac{D}{2}-1\right\vert+ \left\vert \frac{D}{2}\right\rangle \left\langle \frac{D}{2}\right\vert\right) 
\hat{G}_{D}^{\dagger}
\end{equation}

Fig. 2(a) shows the coarse grained current as a function 
of time with Hilbert space dimension $D=32$. We have plotted the results corresponding 
to its maximum negative and positive asymptotic values, for which 
$D_{1}=17$ and $D_{1}=30$ respectively ($s=D_{1}/D\in[0.5,1)$).
It is worth mentioning that $J$ rapidly reaches a stationary behavior (at about 
$t=50$) with fluctuations centered around a finite non-zero value as it is 
illustrated by the smoothed current in Fig. \ref{fig:Jdet32} (solid lines). 
This shows that no unbounded acceleration is present, so there is a rectification 
of transport rather than an effective force in this purely quantum ratchet. 
The classical current for analog initial conditions is also displayed, being zero at all times. 
In Fig. 2(b) we show the probability distribution $P(m,t)$ and its classical version.

In Fig. 3(a) we can see the averaged (asymptotic) coarse grained current 
$\langle J\rangle$ as a function of the Hilbert space dimension $D$ for $D_{1}=D-2$ 
and $D_{1}=D/2-1$ (in order to make this 
averages we have taken the values of the current from $t=100$ up to 
$t=450$). These cases are those at which $J$ 
approximately reaches its maxima in absolute value (positive and negative current, 
respectively). We observe a smooth dependence of this quantity on $D$ (apart from 
small fluctuations), showing that 
the effect is generic. In fact, this situation is 
completely different from previous cases \cite{Lund,Poletti}. In these works 
a purely quantum current was found only at resonant values of a modified KR. 

\begin{figure}[htp]
\centerline{\epsfxsize=7.2cm\epsffile{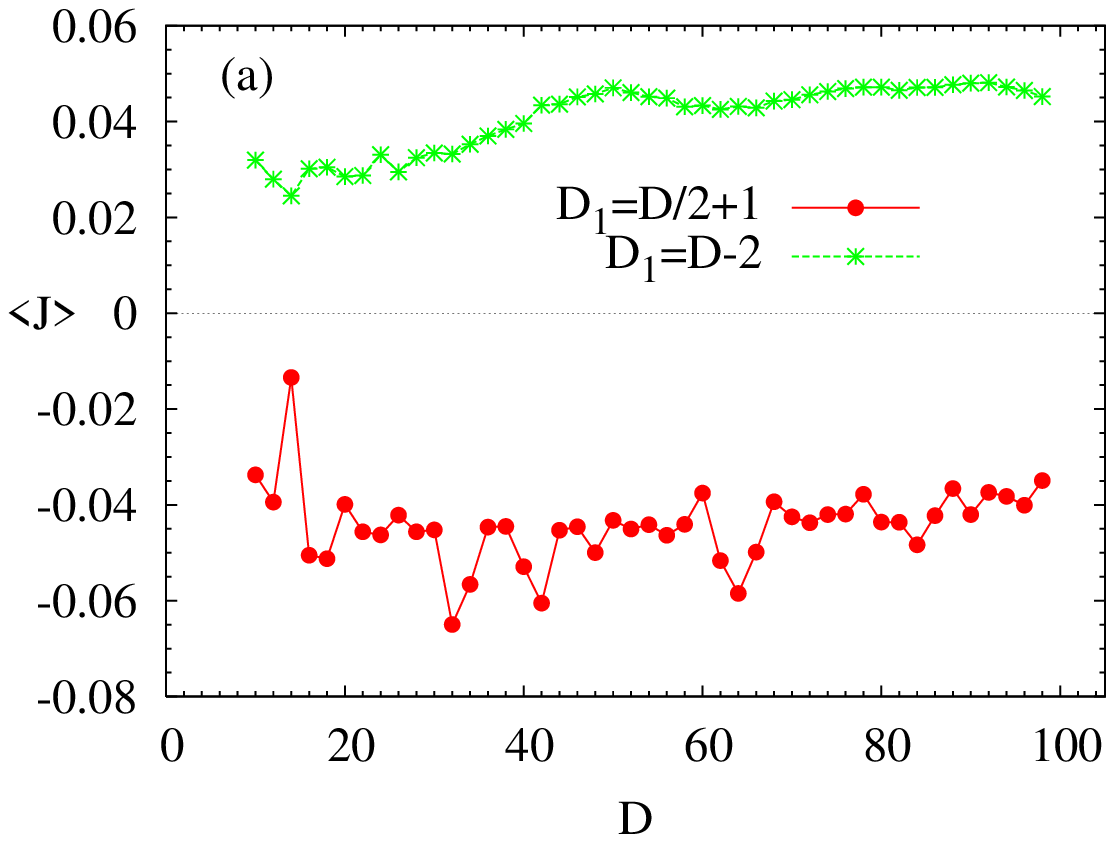}}
\centerline{\epsfxsize=7.2cm\epsffile{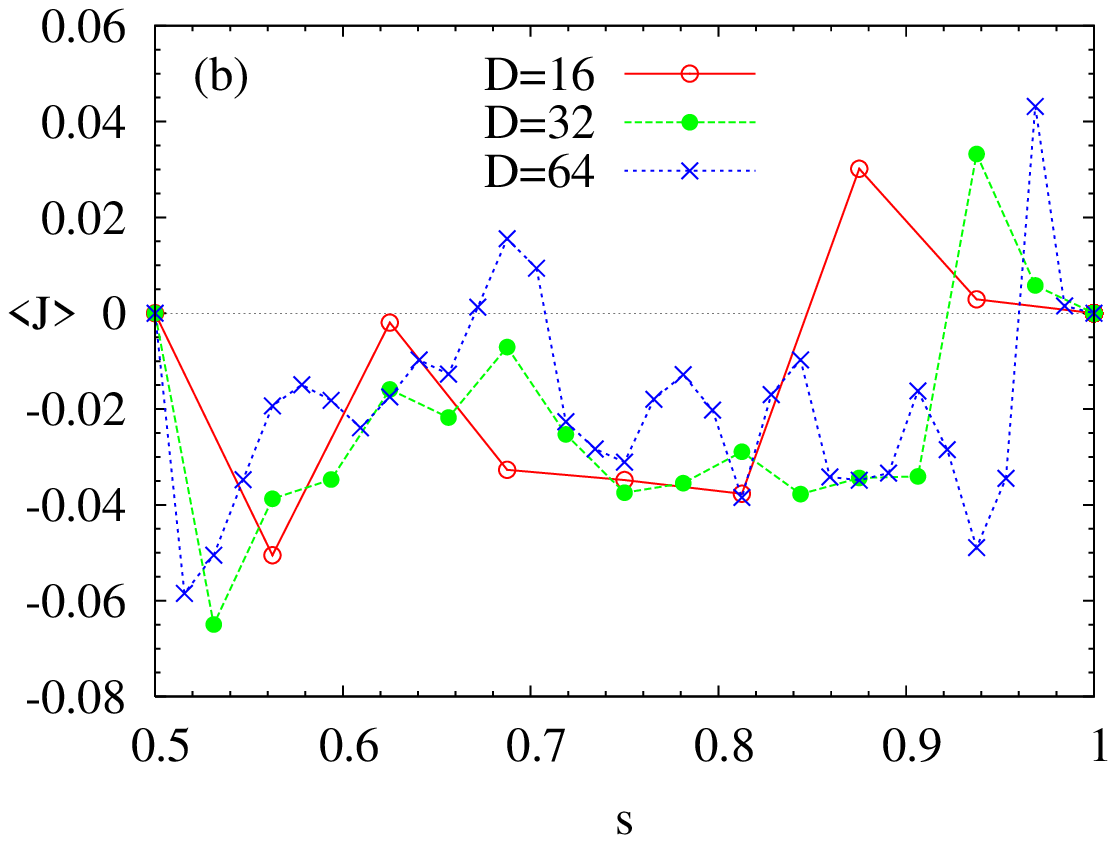}}
\caption{(Color online) Averaged coarse grained current 
$\langle J\rangle$: (a) for different 
values of the dimension of the Hilbert space $(D)$ of the QBM with fixed 
$D_{1}=D/2+1$ and $D_{1}=D-2$; and (b) for the rational values of 
$s=D_{1}/D\in[0.5;1)$ with $D=16,32,64$. The average of $J$ is computed up to 
$450$ iterations of the different maps beginning with $t=100$ (where the current 
approximately reaches its asymptotic behavior).}
\label{fig:JdeD}
\end{figure}

In Fig.\ref{fig:JdeD}(b) we show 
the averaged current vs $s$ for different fixed values of $D$ with $s\geq0.5$. 
We do not show values for $s < 0.5$  
since $J$ turns out to be an odd function of $s$ around 
$s=0.5$ ($\langle J_{s}\rangle=-\langle J_{1-s}\rangle$). 
In fact, if we apply the symmetry transformation $S_I$ to Eq. \ref{qevolution}, 
and then trace out the internal degrees of freedom inside of each cell we 
obtain that $P_s(m,t)=P_{1-s}(-m,t)$ for all $t$. This result is valid for 
any initial $\rho_{\text{QBM}}$ symmetrical under $S_I$. Therefore, the symmetry 
of the current becomes clear (details will be presented 
elsewhere \cite{ErmannF}). As a consequence, 
there is a simple way to obtain current inversion in our system. 

\section{Conclusions}

In summary, we have found a finite asymptotic current in a modified
and unbiased quantum multibaker map; this current has no classical 
counterpart. Moreover, we can invert its direction by exploiting 
a symmetry with respect to the $S_I$ transformation. 
This is much in the same spirit as the corresponding current inversions 
found in the literature \cite{Carlo1,Carlo2}. 

The behavior of our system is rather similar 
to what happens in quantum walks. Different combinations 
of the initial states and/or the unitary operators associated to 
the quantum coin produce a different bias in the final distribution of 
probabilities  \cite{Bach}. 
We believe that this property, that 
is of pure quantum origin is the reason for finding a net transport 
in our system. In fact, we think that the existence of a net current is 
due exclusively to interference effects which persists in the 
$D\rightarrow \infty$ limit. To obtain the classical behavior, i.e., 
a vanishing current, some noise has to be introduced, just as in the case 
of quantum walk \cite{leo1}.

We underline that this is a completely different case 
than that explained in \cite{quantumTheo2}, where the quantum 
current appearance is explained by means of the desymmetrization 
of Floquet states, fact that can be directly related to the corresponding 
classical properties of the system. 
A detailed study of this will be presented elsewhere 
\cite{ErmannF}. Finally, we consider that the features of 
this map can be exploited to design generic Hamiltonian systems that 
behave in a similar way. In that case, there will be the possibility 
to implement them in cold atoms experiments.

\begin{acknowledgments}
Partial support by ANPCyT, CONICET and UNESCO/IBSP Project 3-BR-06 is
gratefully acknowledged.
\end{acknowledgments}



\begin{thebibliography}{99}

\bibitem{Review} 
R. P. Feynmann, {\it Lectures on Physics}, {\bf Vol. 1}, 
(Addison-Wesley,  Reading, MA, 1963); 
R.D. Astumian and P. H\"anggi, Physics Today {\bf 55}, No.11 33 (2002); 
P. Reimann, Phys. Rep. {\bf 361}, 57 (2002);
P. Reimann and P. H\"anggi, Appl. Phys. A {\bf 75}, 169 (2002).

\bibitem{Julicher}
F. J\"ulicher, A. Ajdari, and J. Prost,
Rev. Mod. Phys. {\bf 69}, 1269 (1997).
%

\bibitem{Creffield}
C.E. Creffield and T.S. Monteiro, Phys. Rev. Lett. {\bf 96}, 210403 (2006); 
P.H. Jones {\em et al.}, Phys. Rev. Lett. {\bf 98} 073002 (2007).


%

\bibitem{PolettiR} 
J. Gong, D. Poletti and P. Hanggi, Phys. Rev. A 
{\bf 75}, 033602 (2007); 

\bibitem{CAttractors} P. Jung, J. G. Kissner and P. Hanggi, 
Phys. Rev. Lett. {\bf 76}, 3436 (1996);
J.L. Mateos, Phys. Rev. Lett. {\bf 84}, 258 (2000).


\bibitem{Carlo1} G.G. Carlo, G. Benenti,
G. Casati and D.L. Shepelyansky
Phys. Rev. Lett. {\bf 94}, 164101 (2005).



\bibitem{Flach1} S. Flach, O. Yevtushenko, and Y. Zolotaryuk,
Phys. Rev. Lett. {\bf 84}, 2358 (2000);
S. Denisov and S. Flach, Phys. Rev. E {\bf 64}, 056236 (2001);
S. Denisov {\em et al.}, Physica D {\bf 170}, 131 (2002).


\bibitem{Schanz} H. Schanz, M.F. Otto, R. Ketzmerick, and T. Dittrich,
Phys. Rev. Lett. {\bf 87}, 070601 (2001);
H. Schanz, T. Dittrich, and R. Ketzmerick, 
Phys. Rev. E {\bf 71}, 026228 (2005).


\bibitem{DenisovOrig} 
S. Denisov, J. Klafter, and M. Urbakh, Phys. Rev. E {\bf 66}, 046203 (2002).


\bibitem{quantumTheo} 
T.S. Monteiro, P.A. Dando, N.A.C. Hutchings, and M.R. Isherwood,
Phys. Rev. Lett. {\bf 89}, 194102 (2002);
T. Jonckheere, M.R. Isherwood, and T.S. Monteiro,
Phys. Rev. Lett. {\bf 91}, 253003 (2003);


\bibitem{Carlo2} 
G. G. Carlo {\em et al.}, Phys. Rev. A {\bf 74}, 033617 (2006).

\bibitem{quantumExp} C. Mennerat-Robilliard, D. Lucas, S. Guibal, J. Tabosa,
C. Jurczak, J.-Y. Courtois, and G. Grynberg,
Phys. Rev. Lett. {\bf 82}, 851 (1999); 
M. Schiavoni, L. Sanchez-Palencia, F. Renzoni, and G. Grynberg,
Phys. Rev. Lett. {\bf 90}, 094101 (2003).

\bibitem{quantumTheo2} S. Denisov, L. Morales-Molina, and S. Flach, 
cond-mat/0607558 (2006); S. Denisov, L. Morales-Molina, S. Flach, and 
P. H\"anggi, quant-ph/0703169 (2007).


\bibitem{tunneling}P. Reimann, M. Grifoni, P. Hanggi,
Phys. Rev. Lett. {\bf 79}, 10 (1997).

\bibitem{Brumer} 
I. Franco and P. Brumer, Phys. Rev. Lett. {\bf 97}, 
040402 (2006).

\bibitem{Lund} 
E. Lundh and M. Wallin, Phys. Rev. Lett. \textbf{94}, 110603 (2005).


\bibitem{Poletti}  D. Poletti, G. G. Carlo, and B. Li
Phys. Rev. E \textbf{75}, 011102 (2007)


\bibitem{Gong} J. Gong and P. Brumer, Phys. Rev. Lett. 
{\bf 97}, 240602 (2006).

\bibitem{gaspard} 
P. Gaspard, J. Stat. Phys. \textbf{68} (5/6) 673 (1992);
S. Tasaki and P. Gaspard, J. Stat. Phys. \textbf{81}, (5/6) 935-987 (1995).

\bibitem{wojcik} 
D.K. W\'{o}jcik and J.R. Dorfman, Phys. Rev. E \textbf{66}, 036110 (2002);
Phys. Rev. Lett. \textbf{90}, 230602 (2003);
Physica D \textbf{187}, 223 (2004);
D.K.W\'{o}jcik Int. J. Mod. Phys. B \textbf{20}, 1969 (2006).

\bibitem{voros} 
N.L. Balazs and A. Voros, Ann. Phys. \textbf{190}, 1 (1989).

\bibitem{saraceno} 
M. Saraceno, Ann. Phys. \textbf{199}, 37 (1990).

\bibitem{schackcaves} 
R. Schack and M.C. Caves, AAECC {\bf 10}, 305 (2000).

\bibitem{leo2} 
L. Ermann and M. Saraceno, Phys. Rev. E \textbf{74}, 046205 (2006).

\bibitem{Brun}
 T.A. Brun, H.A. Carteret, and A. Ambainis, Phys. Rev. A \textbf{67}, 052317 (2003).

\bibitem{leo1} 
L. Ermann, J.P. Paz, and M. Saraceno, Phys. Rev. A \textbf{73}, 012302 (2006).


\bibitem{Bach} E. Bach, S. Cppersmith, M.P. Goldschen, R. Joynt and J. Watrous, 
J. Comput. Syst. Sci. \textbf{69}, 562 (2004). 


\bibitem{ErmannF} 
L. Ermann, G.G. Carlo and M. Saraceno, in preparation.




\end{thebibliography}
\end{document}